\documentclass[runningheads]{llncs}
\usepackage{graphicx}
\usepackage{tcolorbox}
\usepackage[linesnumbered, ruled, vlined]{algorithm2e}

\SetCommentSty{mycommfont}
\usepackage{float}
\usepackage{amssymb,amsmath}
\usepackage{hyperref}
\hypersetup{
    colorlinks,
    citecolor=blue,
    filecolor=black,
    linkcolor=red,
    urlcolor=green
}
\begin{document}
\title{Envy\mbox{-}free Trip Planning in Group Trip \\ Planning Query Problem \thanks{Both the authors have contributed equally in this work.}}
\author{Mayank Singhal\inst{1} \and
Suman Banerjee \inst{1} }
\authorrunning{Singhal and Banerjee}
\institute{Department of Computer Science and Engineering, \\ Indian Institute of Technology Jammu, Jammu \& Kashmir 181221, India. \\
\email{2018ucs0064@iitjammu.ac.in, suman.banerjee@iitjammu.ac.in}\\
}
\maketitle              % typeset the header of the contribution
\begin{abstract}
In recent times, Group Trip Planning Query (henceforth referred to as GTP Query) is one of the well\mbox{-}studied problems in Spatial Databases. The inputs to the problem are a road network where the vertices represent the Point-of-Interests (mentioned as POIs henceforth) and they are grouped into different categories, edges represent the road segments, and edge weight represents the distance and a group of users along with their source and destination location. This problem asks to return one POI from every category such that the aggregated distance traveled by the group is minimized. As the objective is to minimize the aggregated distance, the existing solution methodologies do not consider the individual distances traveled by the group members. To address this issue, we introduce and study the \textsc{Envy Free Group Trip Planning Query} Problem. Along with the inputs of the GTP Query Problem, in this variant, we also have a threshold distance $D$ such that aggregated distance traveled by the group is minimized and for any member pairs the difference between their individual distance traveled is less than equal to $D$. However, it may so happen that a given $D$ value no such set POIs are found. To tackle this issue, we introduce the surrogate problem \textsc{Envy Free Group Trip Planning Query with Minimum Additional Distance} Problem which asks what is the minimum distance to be added with $D$ to obtain at least one solution. For these  problems, we design efficient solution approaches and experiment with real-world datasets. From the experiments, we observe that the proposed solution approaches lead to less aggregated distance compared to baseline methods with reasonable computational overhead.    
\end{abstract}
\section{Introduction} \label{Sec:Intro}
Due to the recent advances of mobile devices and wireless networks, storing the location information of moving objects become easy. We call them as trajectory datasets and they are freely available in different data repositories \cite{zheng2015trajectory}. A spatial database contains the location information of different POIs of a large geographical area (e.g. city) and provides services to different kinds of user queries \cite{huang2018location}. One of the well\mbox{-}studied problems in the domain of spatial databases is the Group Trip Planning Query \cite{hashem2013group,tabassum2017dynamic,li2021optimal}. The inputs to the problem are a road network where the vertices represent the Point-of-Interests (mentioned as POIs henceforth) and they are grouped into different categories, edges represent the road segments, and edge weight represents the distance and a group of users along with their source and destination location. This problem asks to return one POI from every category such that the aggregated distance traveled by the group is minimized \cite{singhal2022group}. As the objective is to minimize the aggregated distance, the existing solution methodologies do not consider the individual distances traveled by the group members. It may so happen that the individual distance traveled by a group member is much more than the other one. In such case, the previous group member will be envious to the second one and such solutions even may not be accepted in a realistic situation. To tackle this problem, in this paper we introduce the notion of  Envy-free ness in the GTP Query Problem and call this problem as the \textsc{Envy-free Group Trip Planning Query} Problem.
\par In this new variant, along with the inputs of the GTP Query Problem, we also have a threshold distance $D$ and the goal is to find one POI from each category such that the aggregated distance is minimized as well as for any pair of group members, the difference between the individual distances traveled by them is less than equal to $D$. In practical situations, it is desirable to keep the value of $D$ as much low as possible. However, if we choose the value of $D$ too low then there may be no feasible solution at all. To address this issue, we formulate the Envy-free GTP Query with Minimum Additional Distance Problem. Here the question is how much minimum distance to be added with $D$ such that the feasible solution set for the Envy-free GTP Query Problem will be non-empty. In this paper, we study both these problems and, in particular, we make the following contributions in this paper:
\begin{itemize}
\item  We incorporate the concept of Envy-freeness in the GTP Query Problem and introduce the \textsc{Envy-Free Group Trip Planning Query} Problem.
\item To take care of the situation when the feasible solution set of the Envy-Free GTP Query Problem is empty, we study the Envy-Free GTP Query with Minimum Additional Distance Problem.   
\item We design a solution methodology for solving these problems along with a detailed analysis. Furthermore, we also explore the situation when the spatial data is indexed with $\mathcal{R}$-Tree
\item We conduct an extensive set of experiments with three publicly available trajectory datasets to show the effectiveness and efficiency  of the proposed solution approaches. 
\end{itemize}
Rest of the paper of is organized as follows. Section \ref{Sec:BPD} describes the background and defines the problem formally. Section \ref{Sec:Sol} describes the proposed solution methodologies. Section \ref{Sec:EE} contains the experimental evaluation of the proposed solution approaches. Section \ref{Sec:CFD} concludes our study and gives future research directions.

\section{Background and Problem Definition} \label{Sec:BPD}
In this section, we give the background and defines the problem formally. Initially, we start by defining the road network.

\begin{definition}[Road Network]
A road network is denoted by weighted graph $\mathcal{G}(\mathcal{V}, \mathcal{E}, \mathcal{W})$ where the vertex set $\mathcal{V}(\mathcal{G})=\{v_1, v_2, \ldots, v_n\}$ are the set of $n$ POIs and the edge set $\mathcal{E}(\mathcal{G})=\{e_1, e_2, \ldots, e_m\}$ are the $m$ road segments connecting the POIs. The edge weight function $\mathcal{W}$ assigns each edge to a real number which is basically the corresponding distance between the two POIs of the edge; i.e.; $\mathcal{W}:\mathcal{E}(\mathcal{G}) \longrightarrow \mathbb{R}^{+}$. 
\end{definition}
\par We denote the number of vertices and edges of the road network by $n$ and $m$, respectively. For any edge $(v_iv_j) \in \mathcal{E}(\mathcal{G})$, For any two POIs $v_i$ and $v_j$, we denote its weight by $\mathcal{W}(v_i,v_j)$. For any two POIs $v_i$ and $v_j$, $dist(v_i,v_j)$ denotes the shortest path distance between $v_i$ and $v_j$. A path between the POIs $v_i$ and $v_j$, is a sequence of vertices $<v_{i}, v_{i+1}, v_{i+2}, \ldots, v_{j}>$ such that for all $p \in \{i, i+1, \ldots, j-1\}$, $(v_pv_{p+1}) \in \mathcal{E}(G)$. Also, we assume that the road network is connected that means between any two POIs there exist a path in $\mathcal{G}$. For any two POIs $v_i$ and $v_j$ the shortest path distance is denoted by $dist(v_i,v_j)$. Most of the symbols and notations related to graph theory has been taken from \cite{diestel2005graph}.
\par Now, a group of $b$ friends $\mathcal{U}=\{u_1, u_2, \ldots, u_b\}$ wants to plan a trip in the city with the road network $\mathcal{G}$. Their starting and destination locations are $\mathcal{U}_{\mathcal{S}}=\{v^{s}_1, v^{s}_2, \ldots, v^{s}_b\}$ and $\mathcal{U}_{\mathcal{D}}=\{v^{d}_1, v^{d}_2, \ldots, v^{d}_b\}$, respectively. They want to travel from their respective starting locations through different POIs. As an example, they may decide to first meet at a coffee shop, then to watch movie in a movie theater, subsequently in a hotel for lunch, then in a park for gossiping, and finally to a icecream parlour before heading towards their respective destinations. It is quite natural that in a large city there are multiple locations of each category. Now, which one POI from each category to choose such that total aggregated distance traveled is minimized? This problem is formally referred to as the Group Trip Planning Query Problem.

\begin{definition}[Group Trip Planning Query Problem] \label{Def:GTP_Query}
Given a road network $\mathcal{G}(\mathcal{V}, \mathcal{E}, \mathcal{W})$ where the vertices are marked into $k$ different groups $\mathcal{V}_{1}$, $\mathcal{V}_{2}$, $\ldots$, $\mathcal{V}_{k}$, a group of $b$ friends along with their source and destination locations,  this problem asks to return one POI for every category such that the total aggregated distance traveled by all the group members is minimized.  
\end{definition}
Assume that the POIs $v_1$, $v_2$, $\ldots$, $v_k$ constitute the solution for the GTP Query Problem. Now, it can be observed that the total aggregated distance traveled by the group $\mathcal{U}$ is given by the Equation \ref{Eq:Dist_4}.

\begin{equation} \label{Eq:Dist_4}
\mathcal{D}_{\mathcal{U}}= \underset{i \in [b]}{\sum} \ dist(v^{s}_{i},v_1) \ + \ b \cdot \underset{i \in [k-1]}{\sum} \ dist(v_i, v_{i+1}) + \underset{i \in [b]}{\sum} \ dist(v_k,v^{d}_{i})
\end{equation}
For any $i \in \mathbb{Z}^{+}$, $[i]$ denotes the set $\{1, 2, \ldots, i\}$.  The goal here is the $v_i$s such that $\mathcal{D}_{\mathcal{U}}$ is minimized. Hence, mathematically this problem can be expressed as follows.
\begin{center}
$(v^{*}_1, v^{*}_2, \ldots, v^{*}_k)=\underset{v_1 \subseteq \mathcal{V}_{1}, v_2 \subseteq \mathcal{V}_{2}, \ldots, v_k \subseteq \mathcal{V}_{k}}{argmin} \ \mathcal{D}_{\mathcal{U}}$
\end{center}

Now, it can be observed that this problem is only concerned with the minimization of $\mathcal{D}_{\mathcal{U}}$ not the individual distance traveled by a group member. So, it may so happen that as per the recommended POI the distances traveled by only $u_i$ (denoted by $\mathcal{D}_{u_{i}}$) is much more than that of $u_j$. In this situation, $u_i$ may become envious towards $v_j$ which is not at all desirable. This motivates us to introduce and study the Envy-free Group Trip Planning Query Problem which is stated in Definition \ref{Ref:Def_Envy_Free_GTP}.
%\par So, this solution may not be acceptable to $u_1$ as the distance traveled by $u_1$ and $u_2$ has a huge gap. This may cause $u_1$ to feel envy and he may not agree for such a trip. Hence, it is important to study this problem in a setting which returns the POIs such that the difference between the distance traveled by any two group members are within a given threshold distance. We formally call this problem as the \textsc{Envy-Free Group Trip Planning Query} Problem which is stated in Definition \ref{Ref:Def_Envy_Free_GTP}.

\begin{definition} [\textsc{Envy-Free Group Trip Planning Query} Problem] \label{Ref:Def_Envy_Free_GTP}
Given a road network $\mathcal{G}(\mathcal{V}, \mathcal{E}, \mathcal{W})$ where the vertices are marked into $k$ different categories $\mathcal{V}_{1}$, $\mathcal{V}_{2}$, $\ldots$, $\mathcal{V}_{k}$, a group of $b$ friends along with their source and destination locations, and a threshold distance $D$, this problem asks to select one POI from each category such that the aggregated distance is minimized as well as for any pair of members the individual distance traveled is bounded by the given threshold distance.  
\end{definition}
From the computational perspective, this problem can be written as follows:
%\par For any group member $u_i \in \mathcal{U}$, the distance traveled by him is denoted by $\mathcal{D}_{u_i}$. Hence, $\mathcal{D}_{\mathcal{U}}=\underset{u_i \in \mathcal{U}}{\sum} \ \mathcal{D}_{u_i}$. In case of Envy-free GTP Query Problem the goal is to maximize $\mathcal{D}_{\mathcal{U}}$ subject to the constraint $|\mathcal{D}_{u_i} - \mathcal{D}_{u_j}| \leq D$ for all $u_i, u_j \in $. From the algorithmic point of view, this problem can be posed as follows:
\begin{center}
\begin{tcolorbox}[title=\textsc{Envy-free group trip planning query problem}, width=12cm] 
%\underline{\textsc{Budgeted Influence Maximization Problem}} \\
%\vspace*{2 cm} 
\textbf{Input:}  Road Network $\mathcal{G}(\mathcal{V},\mathcal{E},\mathcal{W})$; The Group $\mathcal{U}=\{u_1, u_2, \ldots, u_b\}$ with their source and destination locations respectively, and a threshold distance $D$.
\vspace{0.2 cm}
\\
\textbf{Problem:} Return the POIs $v^{*}_1 \in \mathcal{V}_{1}$, $v^{*}_2 \in \mathcal{V}_{2}$, $\ldots$, $v^{*}_k \in \mathcal{V}_{k}$ such that the the quantity $\mathcal{D}_{\mathcal{U}}$ is minimized and also  for all $u_i,u_j \in \mathcal{U}$, $|\mathcal{D}_{u_i}-\mathcal{D}_{u_j}| \leq \mathcal{D}$. 
\end{tcolorbox}
\end{center} 
Now, its an important question how to choose the value of $D$. It is easy to observe that if we choose a low value of $D$, then it may so happen that we do not get any solution. So, the immediately next problem is the how much minimum distance to be added with $D$ to obtain a solution for the Envy-free GTP Query Problem. Formally, we call this problem as the \textsc{Envy-free GTP Query with Minimum Additional Distance Problem}. It is important to observe that the second problem comes into existence only when the feasible solution set of the first problem becomes empty. Hence, we call the second problem as the surrogate of the first problem.

%\begin{center}
%\begin{tcolorbox}[title=\textsc{Envy-free GTP query with Minimum Additional Distance}, width=12cm] 
%%\underline{\textsc{Budgeted Influence Maximization Problem}} \\
%%\vspace*{2 cm} 
%\textbf{Input:} Set of $b$ friends $\mathcal{U}=\{u_1, u_2, \ldots, u_b\}$; Their source and destination locations $\{u_1^{s}, u_2^{s}, \ldots, u_b^{s}\}$ and $\{u_1^{d}, u_2^{d}, \ldots, u_b^{d}\}$, respectively; Road Network $\mathcal{G}(\mathcal{V},\mathcal{E},\mathcal{W})$, The hashtag set $\mathcal{X}$ and $\Phi(v)$ for all $v \in \mathcal{V}(\mathcal{G})$, a threshold distance $D$.
%\vspace{0.2 cm}
%\\
%\textbf{Problem:} What is the minimum distance to be added with $D$ such that the set of feasible solutions of the Envy-free GTP Query Problem becomes non-empty. 
%\vspace{0.2 cm}
%
%\textbf{Output:} The minimum threshold distance $\epsilon$ and the corresponding POIs $p_1 \in \mathcal{V}_{\mathcal{C}_{1}}$, $p_2 \in \mathcal{V}_{\mathcal{C}_{2}}$, $\ldots$, $p_k \in \mathcal{V}_{\mathcal{C}_{k}}$.
%\end{tcolorbox}
%\end{center} 

\section{Proposed Solution Approaches} \label{Sec:Sol}
In this section, we describe the proposed solution methodologies for the Envy-free GTP Query Problem. Our first approach is based on the exhaustive search method which has been described in the following subsection.
\subsection{Exhaustive Search Method}
In this method, we exhaustively search among all category of POIs to check whether their exist at least one feasible solution or not. If there exists more than one feasible solution then the one with the least aggregated distance will be returned. If the feasible solution set is empty then the Envy-free GTP Query with Minimum Additional Distance problem is solved. It is important to observe that the number of possible combinations of the POIs can be of $\mathcal{O}(\underset{i \in [k]}{\prod} \ n_i)$ where $n_i$ denotes the number of POIs of $i$-th category. The set of all possible POI Combinations is denoted by $\mathcal{Q}$. Now, the worst case of Algorithm \ref{Algo:1} will arise when the value of this quantity is maximum.Lemma \ref{Ref:Lemma_2} tells the criteria of maximizing this quantity. Due to space limitation we are unable to give the proves.
\begin{lemma} \label{Ref:Lemma_2}
Let $f: \mathbb{R}^{k} \longrightarrow \mathbb{R}$ be a function over a $k$-dimensional real space and $f(x_1, x_2, \ldots, x_{k})= \underset{i \in [k]}{\prod} \ x_i$. This function attains its maximum value subject to the constraint $\underset{i \in [k]}{\sum} x_i \leq n$ when  $x_1=x_2= \ldots=x_k=\frac{n}{k}$.
\end{lemma} 
 In particular, Lemma \ref{Ref:Lemma_2} talks about the distribution of the POIs among the categories and it shows that the worst case of Algorithm \ref{Algo:1} will occur when POIs are equally distributed, and in that case the number of combinations will be of $\mathcal{O}(n^{k})$. In realistic situations, no. of categories of POIs is a constant, and hence, in this study also, we consider that $k$ is a constant.

Now, we describe the working principle of Algorithm \ref{Algo:1} maintains one two dimensional array $Route\_Matrix$ which is a two dimensional array of size $(\underset{i \in [k]}{\prod} n_{i}) \times (k+3)$ and its content is as follows:
\begin{itemize}
\item Corresponding to every POI combination (and certainly there can be at most $\underset{i \in [k]}{\prod} n_{i}$ many) there will be a row in the array $Route\_Matrix$.
\item In each row, among $(k+3)$ many cells, the first $k$ many cells contains the POI combination corresponding to that row.
\item $(k+1)$-th cell contains the aggregated distance of the group considering the POI Combinations as solution.
\item $(k+2)$-th entry stores the maximum difference between the individual distances traveled among the all possible group member pairs.
\begin{equation}
d^{\rho}_{max}= \underset{(u_iu_j) \in \binom{\mathcal{U}}{2}}{max} \ |\mathcal{D}^{\rho}_{u_i} - \mathcal{D}^{\rho}_{u_j}|
\end{equation}
Here, $\rho$ denotes arbitrary POI Combination and $\binom{\mathcal{U}}{2}$ denotes the $2$-element subsets of $\mathcal{U}$.
\item $(k+3)$-th entry stores only a bit. If this entry is $1$, then it signifies that the corresponding POI combination ( which is stored in first $k$ cells in the row) is a feasible solution, else it is not a feasible solution.
\end{itemize}

\begin{algorithm}[H]
  \DontPrintSemicolon
  \KwIn{$\text{The Road Network }\mathcal{G}(\mathcal{V}, \mathcal{E}, \mathcal{W})$, Threshold Distance $D$.}
  \KwOut{Optimal}
$Route\_Matrix = Create\_Array( \underset{i \in [k]}{\prod} \ n_{i},k+3,0)$\;
  
  \For{$\text{All }v_1 \in \mathcal{V}_{1}$}{
  \For{$\text{All }v_2 \in \mathcal{V}_{2}$}{
  $\ldots$\;
  \For{$\text{All } v_k \in \mathcal{V}_{k}$}{
  $\text{The current POIs are }\rho=<v_1, v_2, \ldots, v_{k}>$\;
  $DIS=Create\_Array(b,0)$\;
  \For{$\text{All } u \in \mathcal{U}$}{
  $DIS(u)= dist(u^{s}_1,v_1)+ \underset{i \in [k-1]}{\sum} dist(v_i,v_{i+1})+ dist(v_k,u^{d}_1)$\;
  }
  $Aggr\_Dis \longleftarrow \underset{u \in \mathcal{U}}{\sum} \ DIS(u)$\;
  $Route\_Matrix \longleftarrow (<v_1, v_2, \ldots, v_k>;Aggr\_Dis; --;--)$\;
  $Count \leftarrow 0$; $max \longleftarrow -1$\;
  \For{$\text{All }(u_iu_j) \in \binom{\mathcal{U}}{2}$}{
  \If{$|DIS(u_i)-DIS(u_j)| \leq D$}{
  $Count \leftarrow Count+1$\;
  }
  \If{$|DIS(u_i)-DIS(u_j)|> max$}{
  $max \longleftarrow |DIS(u_i)-DIS(u_j)|$\;
  }
  }
  $Route\_Matrix \longleftarrow (--;--; max ;--)$\;
  \If{$Count==\binom{b}{2}$}{
 $Route\_Matrix \longleftarrow (--;--; --;1)$\;
  }
  }
  $\ldots$\;
  }
  }
  $flag \longleftarrow FALSE$\;
  \For{$i=1 \text{ to }|Route\_Matrix|$}{
  \If{$Route\_Matrix[i,k+3]==1$}{
  $flag \longleftarrow TRUE$\;
  $break$\;
  }
  }
  \eIf{$flag == TRUE$}{
  $<v^{*}_1, v^{*}_2, \ldots, v^{*}_{k}>= argmin \ (Route\_Matrix[;*;;])$\;
  $\text{return } <v^{*}_1, v^{*}_2, \ldots, v^{*}_{k}>$\;
  }
  {
  $d \longleftarrow min (Route\_Matrix[;;*;])$\; 
  $\epsilon \longleftarrow d-D$\;
  $\text{return ``No feasible solution. However, } \epsilon \text{ is added to } D \text{ to have a solution}$.\;
  }
  \caption{Construction of the optimal solution of the Envy Free GTP Query Problem from the GTP Query Problem}
  \label{Algo:1}
\end{algorithm}

    Now, for every combination POIs, we perform the following operations. We compute the individual distances traveled by the group members which requires $\mathcal{O}(b+k) \approx \mathcal{O}(b)$ time. Next, we compute the aggregated distance for the current POI Combinations and  stores it along with the current POI combinations in the array $Route\_Matrix$. This step requires $\mathcal{O}(b)$ time. Initializing two variables count and max requires $\mathcal{O}(1)$ time. As there are $b$ many members hence, the number of member pairs will be of $\mathcal{O}(b^{2})$. Now, for each member pair the time requirement for checking the envy-free ness constraint, updating the count, condition checking of the \texttt{if} statement in Line No. $16$, and subsequently updating the max will take $\mathcal{O}(1)$ time. Hence, time requirement from Line No. $13$ to $17$ will take $\mathcal{O}(b^{2})$ time. Inserting max into its the $Route\_Matrix$ array will take $\mathcal{O}(1)$ time. It is important to observe that for one combination of POIs if for all member pairs, the envy-free ness constraint is satisfied then only we can say that the current POI combination is a feasible solution.  We have incorporated this in the following way. We maintain a \texttt{count} variable and initialized it with zero. Now, we check the envy-free ness constraint for every user pairs one by one and we keep on incrementing \texttt{count} by one. A the end, we check whether the value of \texttt{count} is equals to $\binom{b}{2}$ or not. If it is so, then it clearly signifies that for all the member pairs, the envy-free ness constraint is satisfied. This implies that the current POI combination is a feasible solution for the Envy-free GTP Query Problem, and hence, $(k+3)$-th entry of the current row of $Route\_Matrix$ array is set to $1$. Now, we can easily observe that for one POI combination, time requirement is of $\mathcal{O}(b^{2})$. Hence, total time requirement from Line No. $2$ to $21$ is of $\mathcal{O}(n^{k}b^{2})$.
   \par In the $Route\_Matrix$ array, if we find at least one row such that its $(k+3)$-th entry is $1$ then the feasible solution set of the input instance of the Envy-free GTP Query Problem is non-empty. This has been stated from Line No. $22$ to $25$. First, we set a flag to FALSE. Next, we  scan through the rows of the $Route\_Matrix$ array, and if we find one row whose $(k+3)$-th entry is $1$ then we update the flag to TRUE and breaking out from the loop. It is easy to observe that the time requirement for this step is of $\mathcal{O}(n^{k})$. After this step if the flag is TRUE that means the feasible solution set is non-empty. In that case we need to only scan the  $(k+1)$\mbox{-}th column of those rows of the $Route\_Matrix$ array 
 whose $(k+3)$-th entry is $1$ and returns the POI Combination corresponding to the minimum value of the $(k+1)$\mbox{-}th column. It is easy to observe that the time requirement for this step is of $\mathcal{O}(n^{k})$ time. If the flag remains FALSE after Line No. $26$ then there does not exist any feasible solution for the given instance of the Envy-free GTP Query Problem. In that case, as mentioned previously, Envy-free GTP Query Problem with Minimum Additional Distance needs to be solved. For this, we find out minimum value among $(k+2)$-th entries of all the rows of the $Route\_Matrix$ array.  Assume that this minimum value is $d$. Mathematically, $d$ can be expressed as follows:
 \begin{equation} \label{Eq:3}
 d= \underset{<v_1, v_2, \ldots, v_k> \in  \mathcal{Q}}{min} \ \   \underset{(u_iu_j) \in \binom{\mathcal{U}}{2}}{max} \ |\mathcal{D}^{<v_1, v_2, \ldots, v_k>}_{u_i} - \mathcal{D}^{<v_1, v_2, \ldots, v_k>}_{u_j}|
 \end{equation}

 Without loss of generality, assume that corresponding to one of the minimum values (if there are multiple minimum values) the POI Combination is $<v^{'}_{1}, v^{'}_{2}, \ldots, v^{'}_{k}>$. Then, if we add $(d-D)$ (represented by $\epsilon$ in Algorithm \ref{Algo:1}) to $D$ then we are going to have at least one feasible solution  (and it will be optimal solution as well). Also, as we are considering the minimum value among all the entries of the $(k+2)$-th entries, so $\epsilon$ is the minimum distance to be added to have at least one solution. It is easy to see that this step also requires $\mathcal{O}(n^{k})$ time. Hence, the total time requirement of Algorithm \ref{Algo:1} is of $\mathcal{O}(n^{k} \cdot b^{2} + n^{k} + n^{k})= \mathcal{O}(n^{k} \cdot b^{2})$. The extra space consumed by Algorithm \ref{Algo:1} is to store the $2$d-Array  $Route\_Matrix$ which will take $\mathcal{O}((k+3) \cdot n^{k})=\mathcal{O}(n^{k})$ space and to store the array $DIS$ which will take $\mathcal{O}(b)$ space. Hence, the total space requirement is of $\mathcal{O}(n^{k}+b)$.
%  \par To check whether the feasible solution set is empty or not, we need only $\mathcal{O}(1)$ time. In the worst case, all the POI combinations may be in the feasible solution set, and the number is of $\mathcal{O}(n^{k})$. Now, corresponding to the every POI combination, the table $Route\_Matrix$ contains the aggregated distance. So, the final task is to POI combination corresponding to the minimum aggregated distance is found out and returned as the optimal solution of the Envy-free GTP Query Problem. This step requires $\mathcal{O}(n^{k})$ time. If the feasible solution set is empty then this message is printed and this requires $\mathcal{O}(1)$ time. Hence, it is easy to observe that the total time requirement by Algorithm \ref{Algo:1} is of $\mathcal{O}(n^{k} \cdot b^{2}+ n^{k}) = \mathcal{O}(n^{k} \cdot b^{2})$. Extra space consumed by this algorithm is to store both the data structures $Fea\_Sol\_Set$ and $Route\_Matrix$ which consumes $\mathcal{O}(n^{k})$ and $\mathcal{O}(k \cdot n^{k})$. Considering $k$ is a constant, total space  requirement of Algorithm \ref{Algo:1} is of $\mathcal{O}(n^{k})$. Hence, Theorem \ref{Th:1} holds.
  
  \begin{theorem} \label{Th:1}
  Total time and space requirement of Algorithm \ref{Algo:1} is of  $\mathcal{O}(n^{k} \cdot b^{2})$ and $\mathcal{O}(n^{k}+b)$, respectively.
    \end{theorem}
Now, we state the correctness statement of Algorithm \ref{Algo:1} in Theorem \ref{Th:2}. Due to the space constraint, we are unable to provide its proof.

  \begin{theorem} \label{Th:2}
  Given an Envy-free GTP Query Problem instance if its feasible solution set is non-empty, then Algorithm \ref{Algo:1} correctly returns an optimal solution.  
  \end{theorem}
 Though this approach is easy to understand and simple to implement, the key problem is excessive number of POI Combinations and hence it takes huge computational time.   
 \subsection{Solving the Problem Approximately} 
  To solve our problem approximately, we use the Group Nearest Neighbor (GNN) Search Approach. Given the source location of the group members $\{u^{s}_1, u^{s}_2, \ldots, u^{s}_b\}$ among the POIs in the set $\mathcal{V}_{1}$, one which is the GNN of the POIs $\{u^{s}_1, u^{s}_2, \ldots, u^{s}_b\}$ is returned as the POI (say, in this case $v^{''}_1$) of the first category. Next, among the POIs in $\mathcal{V}_{2}$, one with the nearest neighbor of $v^{''}_1$ is returned as the POI for the second category (say, in this case $v^{''}_2$). Subsequently, among the POIs in $\mathcal{V}_{3}$, one with the nearest neighbor of $v^{''}_2$ is returned as the POI for the third category and so on. This process is continued to search the POIs upto the $(k-1)$-th category. Again, for the POI of $k$-th category again we use GNN  Query approach. So the POI for the $k$-th category $v^{''}_k$ is basically among the POIs of $\mathcal{V}_{k}$ which is GNN of $\{u^{d}_1, u^{d}_2, \ldots, u^{d}_b\}$. It is easy to observe that the number of GNN Queries and NN Queries required are $2$ and $(k-2)$, respectively. The entire process can be  expressed as the following set of equations.

  \begin{center}
  $v^{''}_1= GNN(\{u^{s}_1, u^{s}_2, \ldots, u^{s}_b\}, \mathcal{V}_{1})$\\
  \vspace{0.2 cm}
  $v^{''}_2= NN(v^{''}_1, \mathcal{V}_{2})$\\
  \vspace{0.2 cm}
  $v^{''}_3= NN(v^{''}_2, \mathcal{V}_{3})$\\
  \vspace{0.1 cm}
  $\ldots$\\
  \vspace{0.1 cm}
  $\ldots$\\
  \vspace{0.1 cm}
  $v^{''}_{k-1}=NN(v^{''}_{k-2}, \mathcal{V}_{k-1})$\\
  \vspace{0.2 cm}
  $v^{''}_k= GNN(\{u^{d}_1, u^{d}_2, \ldots, u^{d}_b\}, \mathcal{V}_{k})$
  \end{center}
As we mentioned that between every POIs, the shortest path distance has been computed and stored in a matrix. Now, to analyze this methodology we need to find out how many table loop ups are required. It is easy to observe that for finding $1$-st and $k$-th POI using GNN Queries the number of look ups will be of $\mathcal{O}(b(n_1+n_k))$, and for finding the POIs for $2$-nd onwards upto the $(k-1)$-th category is of $\mathcal{O}(\underset{j \in \{2,3, \ldots, k-1\}}{\sum} n_j)$. As the number of POIs in the road network is $n$, hence the quantity $\mathcal{O}(\underset{j \in \{2,3, \ldots, k-1\}}{\sum} n_j)$ is upper bounded by $\mathcal{O}(n)$. Hence, the total time requirement  to solve the Envy-free Trip Planning Problem approximately is of $\mathcal{O}(b \cdot (n_1+n_k)+n)$. It is easy to follow that this methodology can be implemented in $\mathcal{O}(1)$ space. Hence, Theorem \ref{Th:3} holds.

\begin{theorem} \label{Th:3}
To solve the Envy-free Trip Planning Query Problem approximately using GNN Query-based Approach, time and space requirement is of $\mathcal{O}(b \cdot (n_1+n_k)+n)$ and $\mathcal{O}(1)$, respectively. 
\end{theorem}
 
\section{Experimental Evaluation} \label{Sec:EE}
\begin{figure*}[!ht]
\centering
\begin{tabular}{cc}
\includegraphics[scale=0.2]{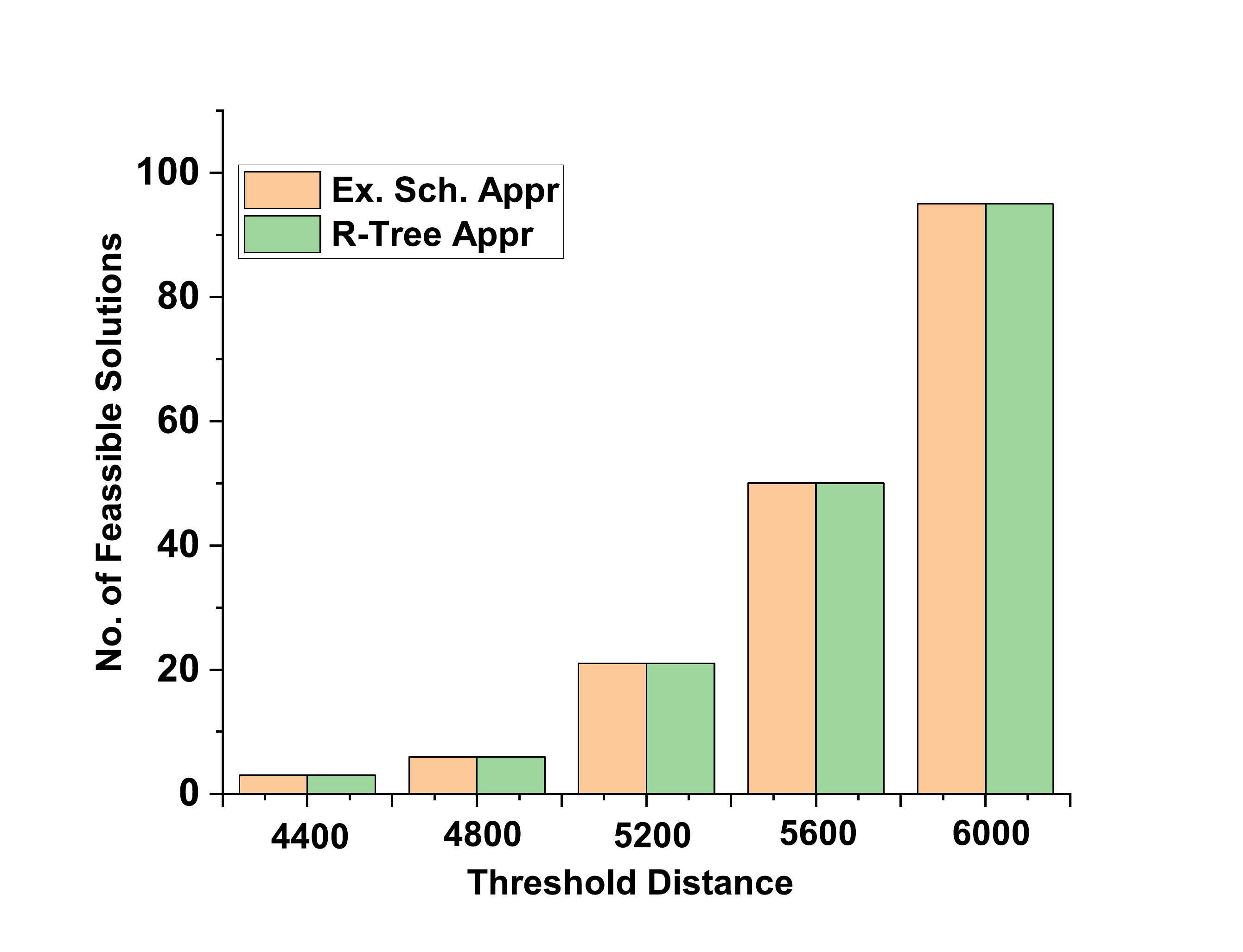} & \includegraphics[scale=0.2]{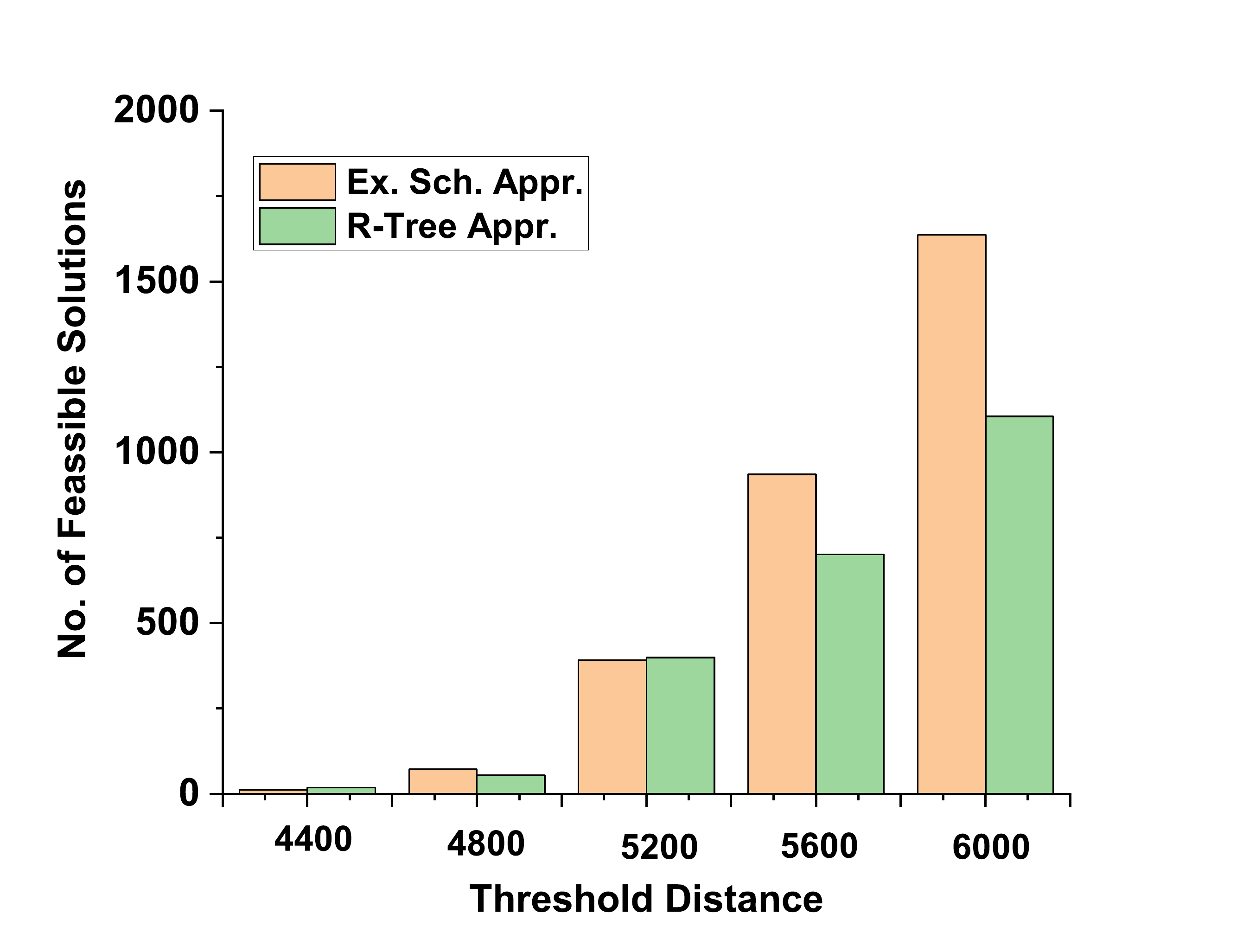} \\
(a) Europe Road Network  & (b) Minnesota Road Network  \\
\includegraphics[scale=0.2]{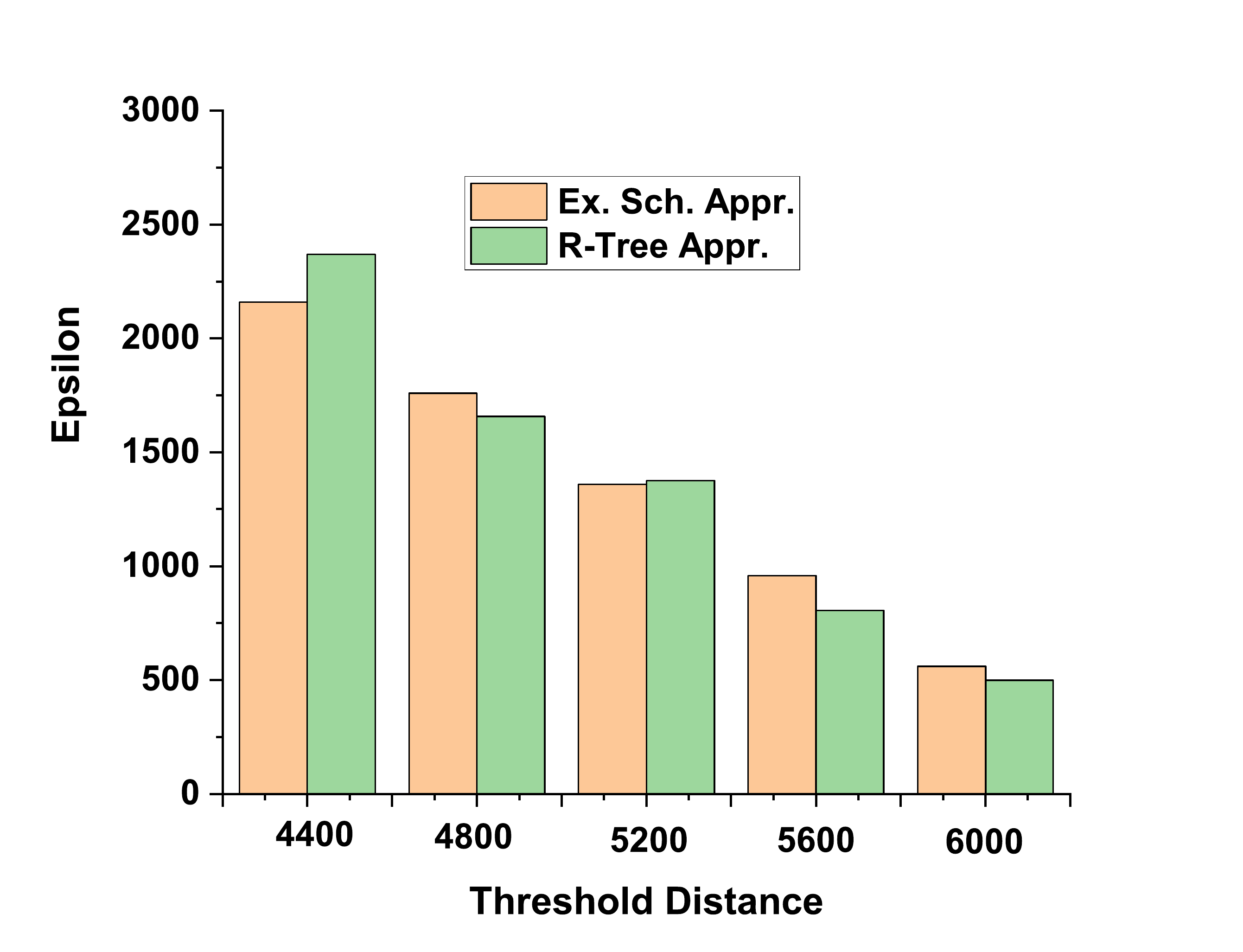} & \includegraphics[scale=0.2]{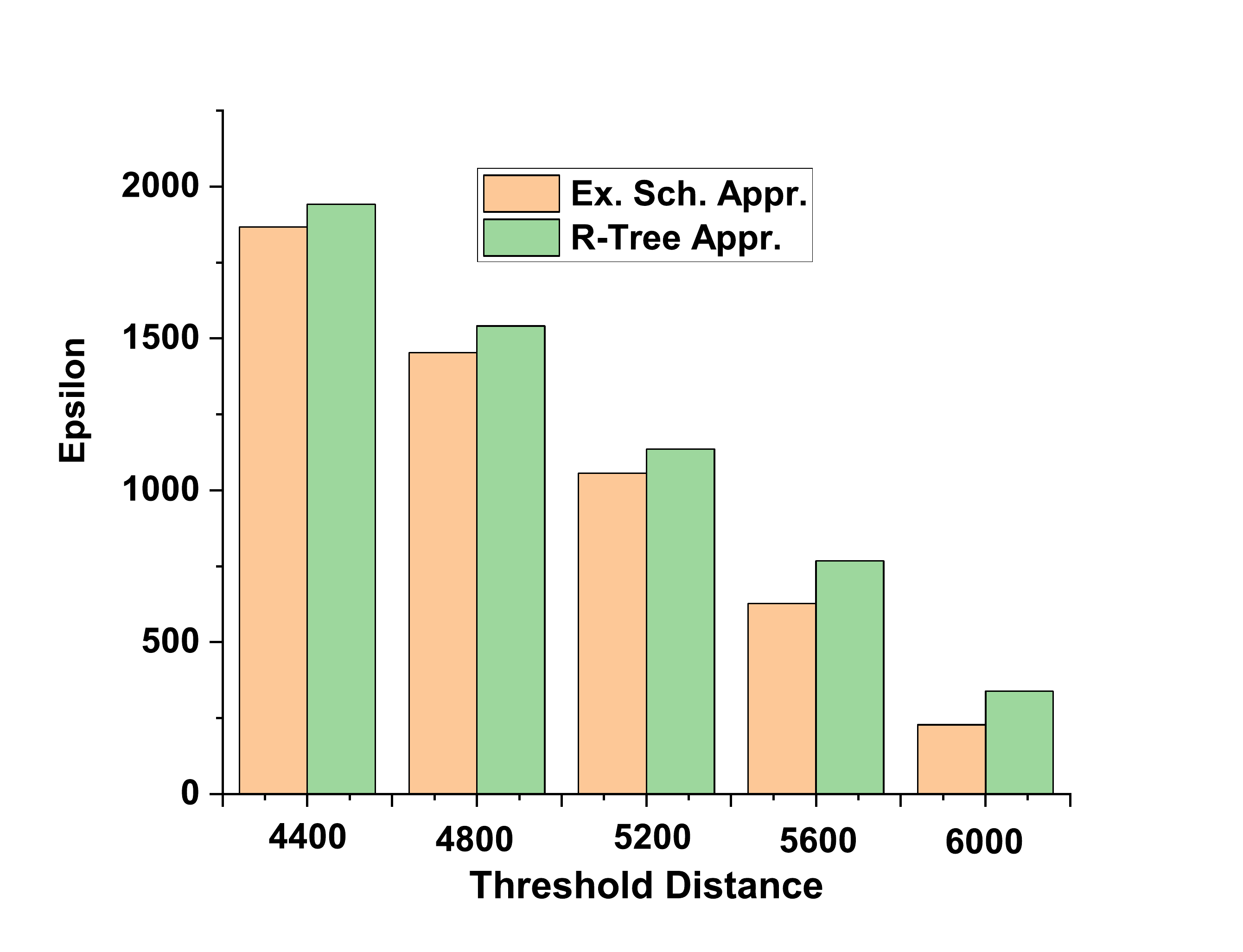} \\

(c) Europe Road Network & (d) Minnesota Road Network  \\
\includegraphics[scale=0.2]{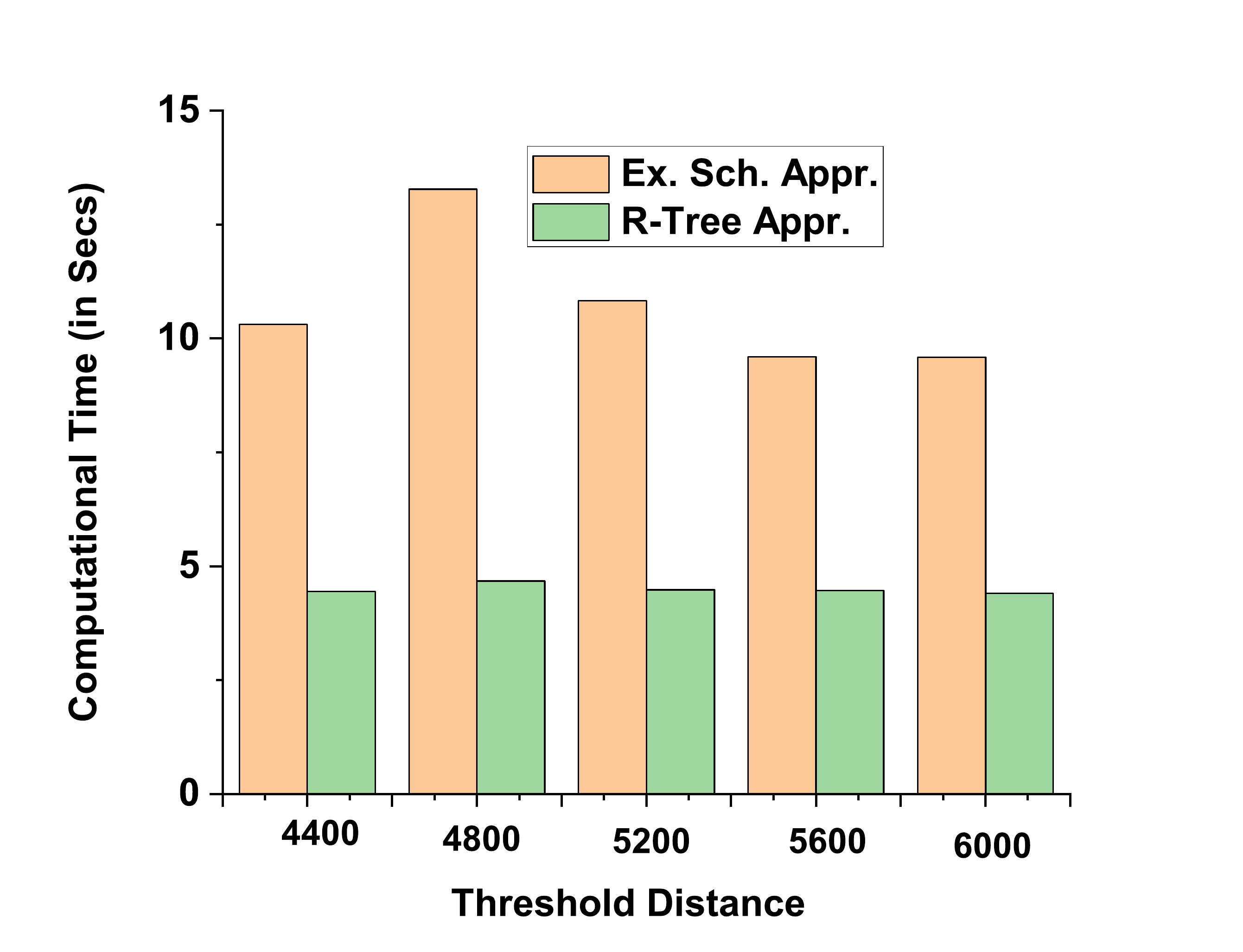} & \includegraphics[scale=0.2]{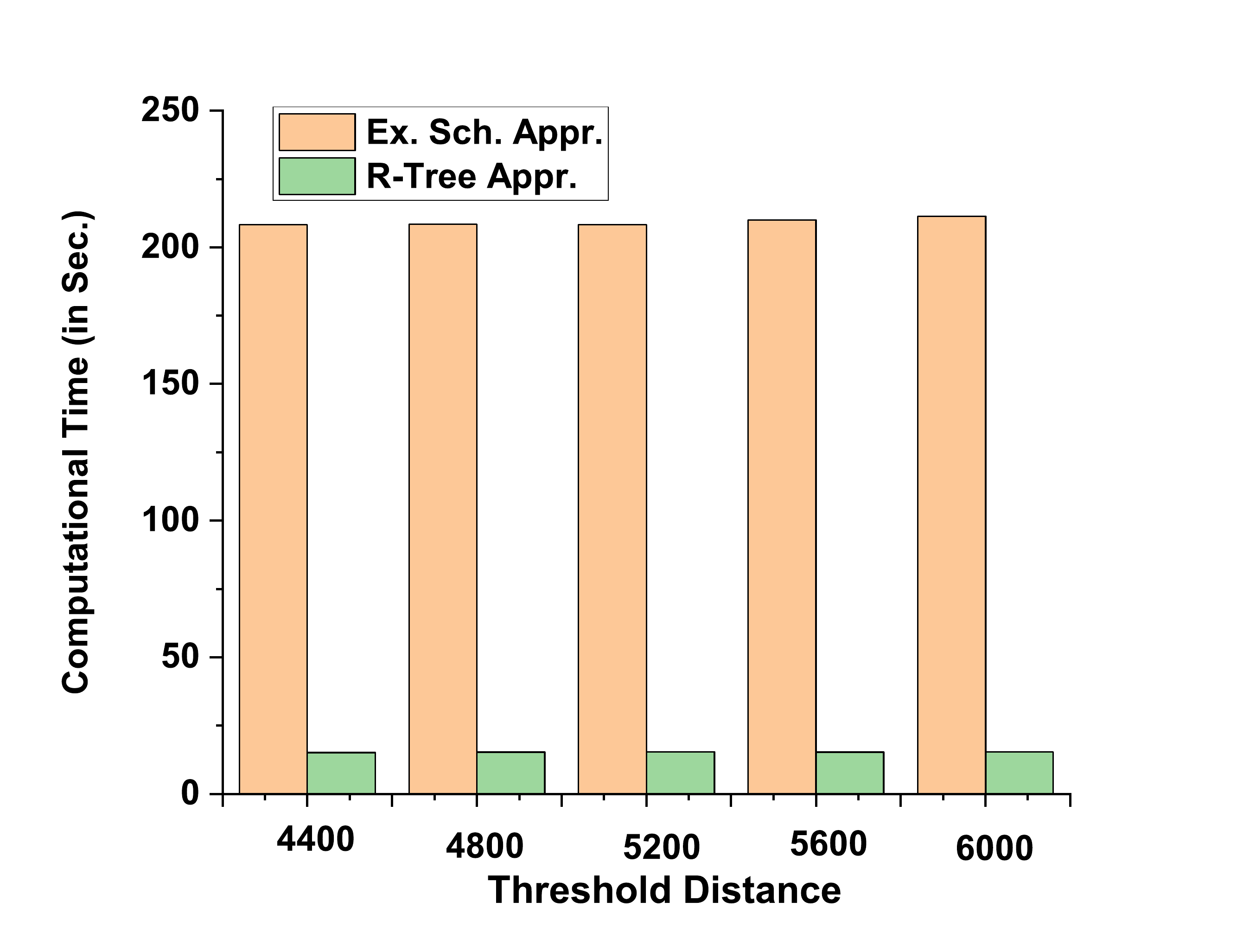} \\
(e) Europe Road Network & (f) Minnesota Road Network  \\
\end{tabular}
\caption{(Group Size, No. of POIs) Vs. Aggregated Distance Traveled for all the methods on three datasets}
\label{Fig:Distance}
\end{figure*}

In this section we describe the experimental evaluations of the proposed methodologies. Initially, we start by describing the datasets.
\paragraph{\textbf{Datasets}} In this study, we use the following two datasets: Europe Road Network and Minnesota Road Network. Both these datasets  are publicly available and have been downloaded from \url{https://networkrepository.com/road.php} \cite{rossi2015network}. The basic statistics of the datasets have been mentioned in Table \ref{tab:Dataset}.
\begin{table}
\centering
  \caption{Basic Statistics of the Datasets}
  \label{tab:Dataset}
  \begin{tabular}{ccccc}
    \hline
    Dataset Name & $n$ & $m$ & Density & Avg. Degree \\
    \hline
    \textbf{Europe Road Network} & 1.2K & 1.4K & 0.00205794 & 2 \\
    \textbf{Minnesota Road Network} & 2.6K & 3.3K & 0.000946754 & 2 \\
  \hline
\end{tabular}
\end{table}
\paragraph{\textbf{Goals of Our Experiments}} In this study, we are interested to answer the following three research questions:
\begin{itemize}
\item With the increase of $D$, how the number of feasible solutions is changing?
\item In case of empty feasible solution set of the Envy-free GTP Query problem is empty, how the $\epsilon$ value is changing with the change of $D$ value and number of categories.
\item How much fast the approximate solution approach is compared to the exhaustive search approach and what is the additional distance incurred for that. 
\end{itemize} 
\paragraph{\textbf{Experiment Setup}} In our study, we fix the group size to be $12$ and we consider the number of categories as $2$, $3$, and $4$. For $D$, we start with the value $4400$ and go upto $6000$ with an increment of $400$ every time.
\paragraph{\textbf{Experimental Results with Discussions}} In this section, we describe our experimental results. Figure \ref{Fig:Distance} shows three different plots for each dataset. The first one is $D$ Vs. No. of feasible solutions. The second one is the $D$ Vs. $\epsilon$, and the third one is No. of categories Vs. Computational Time. From the figure, we observe that for all the datasets when the number of categories are fixed and the threshold distance $D$ is increased then the number of feasible solutions are also increased. As an example, for the Oldenburg Road Network dataset when the number of categories are $2$ and threshold distance is increased from $4400$ to $6000$, the number of feasible solutions are increased from $3$ to $95$. This is due to the following reason. As the threshold distance is increased, then it is natural that more number of POI combinations will satisfy the Envy-free ness Constraint. Also, for the same threshold distance, when the number of categories are increased then also the number of feasible solutions are increasing. As an example, for the Oldenburg Road Network when the threshold distance is $4400$ and the number of categories are increased from $2$ to $4$, then the number of feasible solutions are increased from $3$ to $6084$. When the threshold distance is kept fixed and the number of categories are increased then the number of POI Combinations are also increased. So, it is quite natural that more number of POI Combinations will qualify for the Envy-free ness constraint.

\par Next, we describe our observations related to the second research question. It is easy to observe that for a given problem instance of the Envy-free GTP Query Problem is empty, if its feasible solution set is empty then there exist one value for $d$ (as defined in Equation \ref{Eq:3}). Now, if the value of $D$ is increased then as $\epsilon$ is defined as $(d-D)$, hence $\epsilon$ value will be decreased. In our experiments also, we observe similar phenomenon. As an example, for the Oldenburg Road Network Dataset when the number of categories are fixed and the value of $D$ is increased from $2400$ to $2800$ then the $\epsilon$ value is decreased from $2159.48$ to $1759.67$. In other words, the sum of $D$ and $\epsilon$ is constant. Similar observations are made in case of the San Joqine Road Network dataset also. When the number of categories are $3$ and the value of $D$ is increased from $3600$ to $4000$ the value for $\epsilon$ has been decreased from $687.78$ to $282.56$. For the Oldenburg Road Network dataset, when the $R$-Tree method is used and the number of categories are increased from $2$ to $4$ for the fixed distance threshold $2400$ then the $\epsilon$ value is decreased from $2369.49$ to $2019.48$.
\par Finally, we describe our observation regarding the third research question. It can be observed that the exhaustive search approach scans all possible POI Combinations whether the envy-free ness constraint is satisfied or not. However, in case of approximately solving this problem using Nearest Neighbor Search-based approach we are exploiting the property of the $\mathcal{R}$-Tree and we are not checking all the POI Combinations. Hence, it is expected that the exhaustive search approach will take more time compared to the $\mathcal{R}$-Tree based approach. In our experiments, we observe similar phenomenon in particular for larger problem instance. Hence, from our  experiments we can conclude that if the spatial database is indexed with $\mathcal{R}$-Trees then the problems can be solved in much faster way, particularly when the problem instance size is larger.  
%\subsection{Description of the Datasets}
%\subsection{Experimental Setup}
%\subsection{Baseline Methods}
%\subsection{Goals of the Experimentation}
%\subsection{Experimental Results with Discussions}
\section{Concluding Remarks and Future Research Directions} \label{Sec:CFD}
In this paper, we have introduced and studied two problems related to the trip planning query namely Envey Free Group Trip Planning Query and Envy Free Group Trip Planning Query with Minimum Additional Distance. With one example we have shown that these problems are more relevant compared to their classical version in practice. First, we relate the GTP Query Problem and Envey Free Group Trip Planning Query Problem and show given an optimal solution for the first problem how we can obtain the same for the second one. Subsequently, we propose two efficient algorithms to solve both these problems with detailed analysis and illustration with an example. Experimentation with real-world datasets show that the proposed methodologies can recommend better quality solution compared the baseline methods. In particular, the indexed based approach leads to the best quality solution with reasonable computational time. Now, our study can be extended in the following directions.
 \bibliographystyle{splncs04}
 \bibliography{Paper}
%Author, F.: Article title. Journal \textbf{2}(5), 99--110 (2016)
%Author, F., Author, S.: Title of a proceedings paper. In: Editor,
%F., Editor, S. (eds.) CONFERENCE 2016, LNCS, vol. 9999, pp. 1--13.
%Springer, Heidelberg (2016). \doi{10.10007/1234567890}
%Author, F., Author, S., Author, T.: Book title. 2nd edn. Publisher,
%Author, A.-B.: Contribution title. In: 9th International Proceedings
%on Proceedings, pp. 1--2. Publisher, Location (2010)
%LNCS Homepage, \url{http://www.springer.com/lncs}. Last accessed 4
%\end{thebibliography}
\end{document}